\documentstyle[11pt,newpasp,twoside,epsf]{article}
\def\edcomment#1{\iffalse\marginpar{\raggedright\sl#1\/}\else\relax\fi}
\reversemarginpar

\begin{document}
\title{The Massive Binary Pulsar J1740$-$3052}
 \author{I. H. Stairs}
\affil{Department of Physics and Astronomy, University of British Columbia, 
6224 Agricultural Road, Vancouver, BC V6T 1Z1, Canada}
\author{R. N. Manchester}
\affil{ATNF, CSIRO, P.O. Box 76, Epping NSW 1710, Australia}
\author{A. G. Lyne, M. Kramer}
\affil{University of Manchester,Jodrell Bank Observatory, 
Macclesfield, Cheshire SK11 9DL, UK}
\author{V. M. Kaspi}
\affil{Physics Dept., McGill University, 3600 University St., 
Montreal, Quebec H3A 2T8, Canada}
\author{F. Camilo}
\affil{Columbia Astrophysics Laboratory, 550 W. 120th St., New York, NY 10027, USA}
\author{N. D'Amico}
\affil{Osservatorio Astronomico di Bologna, via Ranzani 1, 40127 Bologna, Italy}

\begin{abstract}
The young pulsar J1740$-$3052 is in an 8-month orbit with a companion of
at least 11 solar masses. We present multifrequency GBT and
Parkes timing observations, and discuss implications for the
nature of the companion.
\end{abstract}

\section{Introduction}

PSR J1740$-$3052 was discovered in 1997 in the Parkes Multibeam Pulsar
Survey (e.g., \cite{mlc+01}), and still leaves a number of fundamental
questions unanswered.  It is a young, unrecycled pulsar ($P =
570$\,ms, $\tau_{\rm c} = 3.5\times10^5$\, kyr) in a 231-day, highly
eccentric orbit about a companion of at least 11 solar masses whose
nature is undetermined.  While there is a late-type star coincident
with the interferometric position of the pulsar, we have argued that
this is probably {\it not} the pulsar companion, and that an early
B-star or even a black hole is more likely (\cite{sml+01}).  
Here we present an up-to-date analysis of ongoing 
multifrequency observations of this pulsar, and explore what these 
data may tell us about the true pulsar companion.

\section{Observations and Timing}

Parkes observations were taken at 2380, 1400 and 660\,MHz, using
multichannel filterbanks and 1-bit digitization.  The 1400\,MHz data
have been taken regularly since mid-1998, while data at the other two
frequencies have been acquired primarily in the weeks leading up to
various periastron epochs.  Some 1400\,MHz data were also acquired
regularly at Jodrell Bank Observatory from mid-1998 through the end of
2000.  Since Sept. 2001, we have been monitoring PSR J1740$-$3052 with
the new 100-m Green Bank Telescope (GBT), using regular 3--4 week
spacing with more frequent sampling in the lead-ups to periastrons.
At each epoch, multifrequency data were acquired with the
``Berkeley-Caltech Pulsar Machine'' (BCPM) flexible filterbank:
typically centre frequencies of 2200, 1400, 1190 and 590\,MHz were
used, with occasional points at 1780 or 820\,MHz.  All profiles were
dedispersed and folded according to the predicted topocentric period.
TOAs were obtained by cross-correlation and fit to a pulse timing
model using {\sc tempo}\footnote{\tt
http://pulsar.princeton.edu/tempo}.  The templates at different
frequencies were aligned by cross-correlation and (for those with
scattering tails) by timing.

\begin{figure}
\plotfiddle{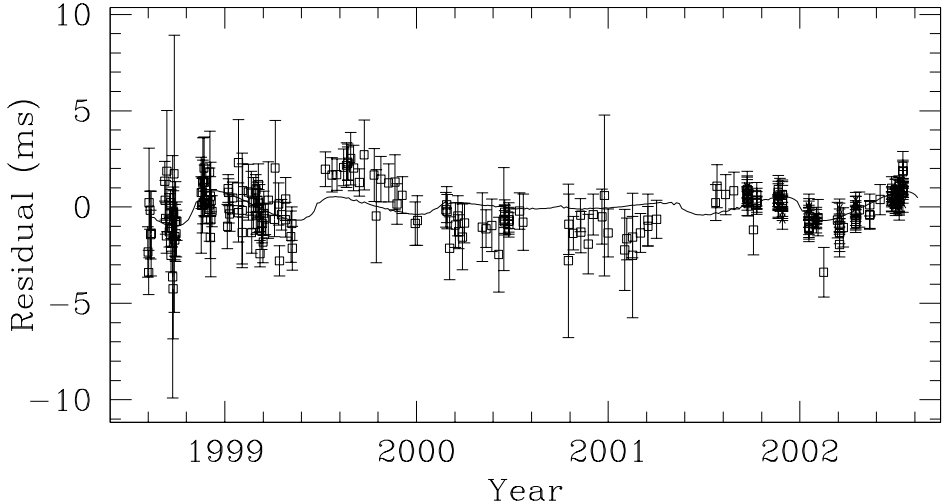}{3.0in}{0}{65}{65}{-200}{-100}
\caption{Preliminary measurement of $\dot x$, the apparent change
  in the size of the orbit from precession induced by the spin mass
  quadrupole of the companion.  The solid line indicates the predicted
  residuals, for an $\dot x$ corresponding to our measured value,
  relative to a solution with constant $x$.  The points are actual
  timing residuals from Parkes, Jodrell Bank and the GBT.\label{fig:xdot}}
\end{figure}

The timing model used was that for pulsar--main-sequence binaries
(\cite{wex98}).  Along with the five standard Keplerian orbital
parameters, two corrections representing secular evolution of the
observed orbit must also be fit.  These are the advance of periastrion
($\dot \omega$, now a 5-$\sigma$ detection), and a preliminary
(4-$\sigma$) measurement of $\dot x$, the apparent change of the
projected semi-major axis of the orbit (see Figure~\ref{fig:xdot}).
The measured $\dot \omega$ is likely due to a mixture of general
relativistic orbital precession and precession due to the spin
quadrupole of the companion star; $\dot x$ can only be interpreted as
due to the companion spin -- the pulsar is distant ($\sim 11$ kpc) and
its proper motion is not measurable for reasonable velocities, so this
cannot affect the observed $\dot x$ (\cite{kop96}).  Intriguingly, the
measured value of $\dot \omega$,
$1.89\pm0.35\times10^{-4\,\circ}$yr$^{-1}$, is very close to the
general relativistic prediction for an edge-on orbit.  It is therefore
important to try to estimate the inclination angle of the system, as
this will determine the spin-quadrupole contribution.

\begin{figure}
\plotfiddle{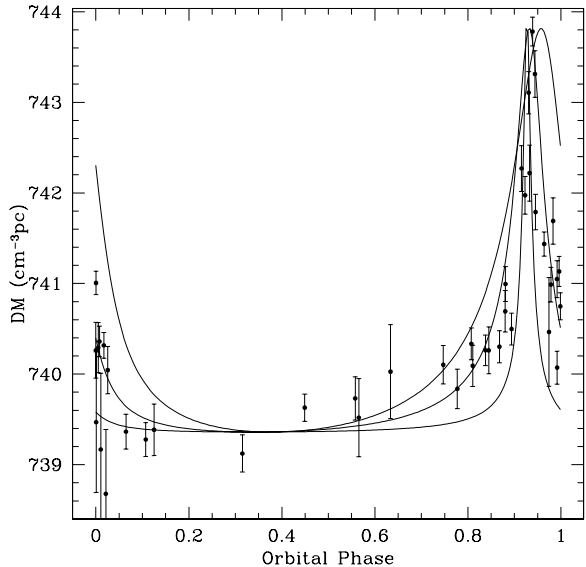}{3.0in}{0}{40}{40}{-130}{-55}
\caption{The points indicate observed values of the pulsar's
  dispersion measure (DM) as a function of orbital phase, from data
  obtained with the GBT and Parkes telescope.  The curves show the
  expected changes in the DM if the pulsar signal passes through a
  spherically symmetric B-star wind, for inclination angles of
  40$^{\circ}$ (top), 70$^{\circ}$ and 85$^{\circ}$ (bottom), with the
  overall magnitude of the DM predictions scaled to match the observed
  values.  The curves are {\it not} fits to the data.\label{fig:dmvar}}
\end{figure}

\section{DM Variations and Wind Model}

The multifrequency GBT and pre-periastron Parkes data allow us to
estimate the dispersion measure (DM) at multiple epochs throughout the
orbit (Figure~\ref{fig:dmvar}).  The most abrupt changes occur in the
month before periastron, as the pulsar swings ``behind'' its
companion.  The measured DM variations are 2--3 orders of magnitude
too small to be accounted for by a K-star wind (\cite{sml+01}).
Instead, we use a simple model of the mass-loss and velocity for a
line-driven B-star wind (see \cite{ktm96} and references therein) and
integrate the total expected DM contribution from the wind along the
line of sight to the pulsar.  We then scale these predictions to match
the range of observed DM variations, and compare the {\it shape} of
the predicted curve to the observed values.  A simple $\chi^2$
analysis indicates that the best inclination angle is around
70$^{\circ}$--75$^{\circ}$ and the required scaling implies a
mass-loss rate $2\times 10^{-9}\,M_{\odot}$/yr and terminal velocity
2200 km/s, which are typical values for a B-star.  The model is
therefore self-consistent.

\section{Implications for Orbital Geometry}

An estimated inclination angle of 70$^{\circ}$ implies a companion
mass of 12.8\,M$_{\odot}$ assuming a neutron-star mass of
1.35\,M$_{\odot}$ (\cite{tc99}).  The spin quadrupole contribution to
$\dot \omega$ is therefore small and likely negative.  Its exact value
depends not only on the rotation rate and internal structure of the
star, but also on the precession phase $\Phi_0$ and the misalignment
$\theta$ of the stellar spin with the orbital angular momentum.  The
spin-induced $\dot x$ depends in the same way on the stellar spin and
structure, and so the ratio of the two measured parameters can be used
to constrain the two unknown angles (\cite{wex98,kbm+96}):
\begin{displaymath}
\frac{\dot \omega\,x}{\dot x} \sin \Phi_0 -\cos \Phi_0 = \frac{1+3\cos2\theta}{2\cot i \sin2\theta},
\end{displaymath}
where $i$ is the inclination angle of the orbit.  Allowing for
2$\sigma$ variation in both $\dot \omega$ and $\dot x$, we find that
$\theta$ is most likely to be in the range
$30^{\circ}$--$75^{\circ}$.  This misalignment is likely due to an
asymmetric kick in the supernova that created the pulsar.  With
ongoing long-term timing, we will refine the measurements of $\dot
\omega$ and $\dot x$ and begin to search for the predicted second
derivative of each parameter, which will lead to further constraints
on the orbital geometry.

\end{document}